\documentclass[reprint,amsmath,amssymb,aps]{revtex4-1}
\pdfoutput=1

\usepackage{graphicx}% Include figure files
\usepackage{dcolumn}% Align table columns on decimal point
\usepackage{bm}% bold math
\begin{document}

\preprint{APS/123-QED}

\title{Detecting large extra dimensions with optomechanical levitated sensors}% Force line breaks with \\
\thanks{}%

\author{Jian Liu}
 \altaffiliation[]{}%Lines break automatically or can be forced with \\
\author{Ka-Di Zhu}%
 \email{zhukadi@sjtu.edu.cn}
\affiliation{Key Laboratory of Artificial Structures and Quantum Control (Ministry of
Education), School of Physics and Astronomy, Shanghai Jiao Tong
University, 800 DongChuan Road, Shanghai 200240, China,
Collaborative Innovation Center of Advanced Microstructures, Nanjing, China
}%

\date{\today}% It is always \today, today,
             %  but any date may be explicitly specified

\begin{abstract}
Numbers of tabletop experiments have made efforts to detect large extra
dimensions for the range from solar system to submillimeter system, but the
direct evidence is still lacking. Here we present a scheme to test the
gravitational law in 4+2 dimensions at microns by using cavity
optomechanical method. We have investigated the probe spectrum for coupled
quantum levitated oscillators in optical cavities. The results show that the
spectral splitting can be obtained once the large extra dimensions present.
Compare to the previous experiment, the sensitivity can be improved by the
using of a specific geometry and a shield mirror to control and suppress the
effect of the Casimir background. The weak frequency splitting can be
optically read by the pump-probe scheme. Thus we can detect the
gravitational deviation in the bulk based ADD model via spectroscopy without
the isoelectronic technique.

\end{abstract}

\pacs{Valid PACS appear here}% PACS, the Physics and Astronomy
                             % Classification Scheme.
%\keywords{Suggested keywords}%Use showkeys class option if keyword
                              %display desired
\maketitle

%\tableofcontents

\section{INTRODUCTION}

Why gravity is so weak compared to the other known forces in nature? This
can be recast by the hierarchy problem, the seeming disparity (16 orders of
magnitude) between the Planck mass and standard model electroweak scale.This
problem can be overcame by adding new dimensions in the large extra
dimension(bulk) model, namely the ADD model, which was first developed by
Arkani-Hamed, Dimopoulos, and Dvali[1-3]. They proposed that for $n=2$, the
extra dimensions could be as large as a millimeter and the measurements of
gravity may observe the transition from $1/r^{2}$ to $1/r^{4}$ Newtonian
gravitation. In addition to this possibility, they found that their model
could describe the hierarchy between the Planck mass and the electroweak
symmetry breaking scale in terms of the large size of the extra dimensions.
Therefore testing general relativity and its Newtonian limit at short
distances has become particularly important in light of recent theoretical
developments[4-13].

On the other hand, significant advances have been witnessed in studying the
characteristic and application of the cavity optomechanical system with
high-Q mechanical nanoresonators(NRs)[14]. Nano- and micromechanical devices
can be coupled to optical cavities directly via radiation pressure leading
to a variety of important properties, such as optical self-focusing[15],
optomechanical entanglement[16] and optomechanically induced transparency
(OMIT)[17,18]. Recently much effort has been directed toward optically
levitating nano- and micro-mechanical oscillators in ultrahigh vacuum such
as nanospheres[19,20], nanodiamonds[21,22], microdisk[23,24], and even the
living organisms[25]. The laser trapped objects has no physical contact to
the environment, leads to ultralow mechanical damping. In the absence of
other noise sources, the quality factors of optical levitated
nano-optomechanics can exceed $10^{12}$ with pressures of $10^{-10}$
Torr[19]. Owing to its exceptional mechanical properties, optically
levitation appears to be an excellent candidate for the NRs in
nanoelectromechanical systems (NEMS), which are of great interest for
applications in fundamental science and precision measurement[26-28].

In the present work, by combining the cavity optomechanics and the ADD
model, we investigate the optomechanical system consisted of coupled quantum
levitated oscillators and cavities. Then we propose a design for
non-Newtonian gravity detection at short range with the levitated sensors.
The results show that the sharp enhanced transparency peaks with
ultra-narrow linewidth can be induced through the coupling between the
optical cavity and the levitated resonator. The contribution of extra
dimensions will be able to show itself clearly on the probe spectrum by
spectral splitting, performing a test of Newton's law at microns. We also
consider the Casimir effect as the main background force noise in the
micro-scale optomechanical system. The constraints on the compactification
distances are deduced as last, the super-resolution can be achieved by using
the oscillators with high Q factor in vacuum. We expect that the proposed
scheme could be applied to probe the large extra dimensions or set a new
upper limit on the hypothetical long-range interactions which naturally
arise in many extensions to the standard model[29,30].

\section{THEORY FRAMEWORK}

A schematic of our setup is sketched in Fig.1(a), where a dielectric
microdisk and a nanosphere are optically levitated in two optical cavities.
The micro- or nano-scale objects is attracted to the anti-node of the field.
The resulting gradient in the optical field provides a sufficiently deep
optical potential well which allows the object to be confined in a number of
possible trapping sites, with precise localization due to the optical
standing wave[25,31,32]. We use a trapping laser to levitate the microdisk
in the left cavity and use the other beam to trap the nanosphere in the
right cavity. Then the left cavity is driven by a strong pump laser and
probed by a weak probe laser. 
The mechanism can be explained as a quantum coupling induced normal mode splitting(NMS)
in the four-wave mixing (FWM)
process. The Fig.1(b) is the energy level description of this process. In
the present paper, the trapped microdisk and nanosphere are treated as
quantum-mechanical harmonic oscillators and their masses are $m_{1}$ and $%
m_{2}$, mechanical frequencies $\omega _{1}$ and $\omega _{2}$, and damping
rates $\gamma _{1}$ and $\gamma _{2}$, respectively. The Hamiltonian can be
regarded as $H_{\omega }=\underset{j=1,2}{\sum }\hbar \omega
_{j}a_{j}^{+}a_{j}$, where $a_{j}^{+}$ and $a_{j}$ are the bosonic creation
and annihilation operators for the two vibrational resonators. We use $%
H_{c}=\hbar \omega _{c}c^{+}c$ to describe the Hamiltonian of the left
cavity mode, here $\omega _{c}$ and $c(c^{+})$ denote the oscillation
frequency and the annihilation (creation) operator of the cavity. The
radiation pressure of the cavity gives rise to the optomechanical coupling $%
H_{g}=-\hbar g_{1}c^{+}c(a_{1}^{+}+a_{1})$, where $g_{1}$ is the
single-photon coupling rate between the microdisk and the left cavity, it
has the typical value of $\sim 2\pi \times 1.2Hz$[31].
\begin{figure*}[tbp]
\includegraphics[width=16cm]{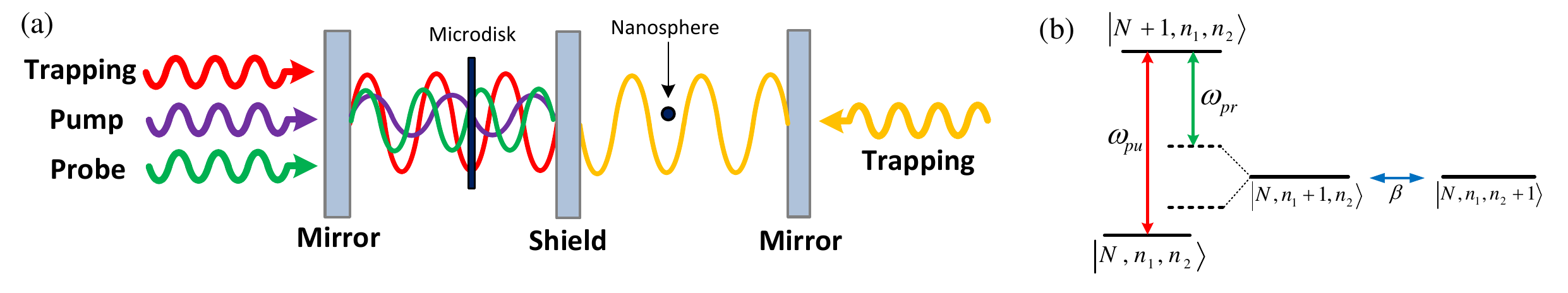}
\caption{(a)Schematic diagram of the double oscillators cavity
optomechanical system. The left cavity is driven by a strong pump laser and
probed by a weak signal laser. The levitated microdisk and nanosphere are
placed at the antinodes of cavity field with a separation of $r_{0}=8\protect%
\mu m$. (b)Schematic of the energy-level diagram for the resonance mode
splitting in the left cavity.}
\end{figure*}

Then we consider the gravity interaction in the system. The ADD theory
establishes an effective Planck scale to coincide with the electroweak scale
by allowing gravity to travel in extra dimensions. It is possible that all
of the particles and fields of the standard model are trapped on this brane,
providing an explanation for why we have never observed more than our three
spatial dimensions. On the other hand, gravity would be able to travel in $%
(4+n)$ dimensions, where $n$ is the number of extra dimensions. According to
this model, the Planck scale is not fundamental but determined by the volume
of the extra dimensions $M_{Pl}^{2}\sim M_{Pl(4+n)}^{2+n}V_{n}$. By this
process in ADD, the hierarchy problem is nullified and a modification of
Newtonian gravity is proposed for the ranges smaller than the
compactification length. When the separation between the masses decreases to
the point where $r\ll R_{\ast }$($R_{\ast }$ denotes the size of the extra
dimensions), the usual inverse square law of gravity changes to a new
power-law[3]
\begin{equation}
V(r)=-\frac{m_{1}m_{2}}{M_{Pl(4+n)}^{n+2}}\frac{1}{r^{n+1}}.
\end{equation}%
The separation can be regarded as $r=r_{0}+x_{1}-x_{2}$, where $x_{1,2}$
denote the displacements of mechanical oscillators from their equilibrium
positions, $r_{0}$ denotes the distance between equilibrium positions of the
disk and sphere. For equal-size extra dimensions and toroidal
compactification, $V_{n}=(2\pi R_{\ast })^{n}$ and the size of the extra
dimensions,
\begin{equation}
R_{\ast }=10^{30/n-17}mm\times (\frac{1TeV}{m_{EW}})^{1+\frac{2}{n}}.
\end{equation}%
Here the electroweak scale $m_{EW}$, also known as the Fermi scale, is the
energy scale around $246GeV$. For $n=1,R_{\ast }\sim 10^{13}cm$ implying
deviations from Newtonian gravity over solar system distances. In the case
of $n=2$, $R_{\ast }\sim 1mm$, it is particularly exciting, since it is the
scale currently being probed by a number of tabletop gravity experiments. In
our scheme the levitated resonators held in trapping potentials are
separated by a distance of $r_{0}=8\mu m$, thus the condition of $r\ll
R_{\ast }$ can be physically achieved. Expanding Eq.(1) in the condition of $%
\left\vert x_{1}\right\vert ,\left\vert x_{2}\right\vert \ll r_{0}$ and
working to the lowest order, we can obtain the term of gravitational
potential for 4+n dimensions as%
\begin{equation}
\begin{split}
V_{int}& \approx -G_{4+n}m_{1}m_{2}[c_{n,0}\frac{1}{r_{0}^{n+1}}-c_{n,1}%
\frac{(x_{1}-x_{2})}{r_{0}^{n+2}} \\
& +c_{n,2}\frac{(x_{1}-x_{2})^{2}}{r_{0}^{n+3}}],
\end{split}%
\end{equation}%
where $G_{4+n}\sim M_{Pl(4+n)}^{-(2+n)}$ is the fundamental gravitational
constant in the full 4+n dimensional spacetime, $C_{n,k}=(-n-1)$\ldots $%
(-n-k)/k!$ is the binomial coefficient. The first term is constant. The
second term represents a steady force does not affect the interactional
dynamics. The term proportional to $x_{1}x_{2}$ represents the lowest-order
coupling between the resonators' motions. In the regime of $\omega
_{c},\omega _{j}\gg g_{j}$, the Hamiltonian of gravitational interaction can
be obtained by quantizing mechanical oscillators within rotating wave
approximation[33]
\begin{equation}
H_{int}=C_{n,2}\frac{G_{4+n}m_{1}m_{2}}{r_{0}^{n+3}}2x_{1}x_{2}\cong \hbar
\beta (a_{1}^{+}a_{2}+a_{1}a_{2}^{+}),
\end{equation}%
and
\begin{equation}
\beta (n)=C_{n,2}\frac{G_{4+n}}{r_{0}^{n+3}}\frac{\sqrt{m_{1}m_{2}}}{\sqrt{%
\omega _{1}\omega _{2}}},
\end{equation}%
The coefficient $\beta $\ can be defined as the bulk induced coupling rate,
which reveals the gravitational strength between two oscillators in 4+n
dimensions.

For smaller separations($\ll 10^{-4}m$), Casimir forces provide the dominant
background force, and we expect the scheme can set the best limits in this
regime. In the original point of view, the Casimir effect is derived from
the change of the total energy of vacuum due to the presence of two plane
perfect reflectors. In this global approach, the Casimir energy is the part $%
H_{casmir}$ of vacuum energy depending on the plate separation $r_{0}$. For $%
r_{0}\gg r_{2}$, the Casimir energy between a sphere and plane takes the
Casimir-Polder form[34], $H_{casmir}=3\hbar c\alpha _{V}/32\pi
^{2}\varepsilon _{0}r_{0}^{4}$. We can manipulate it with the same means and
obtain the Casimir coupling rate%
\begin{equation}
\beta _{casimir}=\frac{3C_{3,2}\hbar c\alpha _{V}}{32\pi ^{2}\varepsilon
_{0}r_{0}^{6}}\frac{1}{\sqrt{m_{1}m_{2}}\sqrt{\omega _{1}\omega _{2}}},
\end{equation}%
where $\alpha _{V}=3\epsilon _{0}V(\epsilon _{2}-1)/(\epsilon _{2}+2)$ is
the electric polarizability, and $V$ is the volume of the nanosphere, $%
\epsilon _{2}$ is the dielectric constant of the nanosphere. Thereby the
total interaction Hamiltonian, including Casimir interaction, can be written
as $H_{int}^{\prime }=\hbar (\beta +\beta _{casimir})\cdot
(a_{1}^{+}a_{2}+a_{1}a_{2}^{+})$.

Based on the discussion above, the whole system can be considered as an
optomechanical system where the microdisk's vibrational mode is coupled to a
single cavity mode with the coupling rates $g_{1}$, meanwhile interacts with
the other levitated oscillator(nanosphere) through the gravitational
strength $\beta $ and the Casimir coupling rate $\beta _{casimir}$. By
applying the pump-probe field, we can obtain the Hamiltonian of the whole
system as%
\begin{equation}
\begin{split}
H& =H_{c}+H_{\omega }+H_{g}+H_{int}^{\prime }-i\hbar \Omega _{pu}(c-c^{+}) \\
& -i\hbar \Omega _{pr}(ce^{i\delta t}-c^{+}e^{-i\delta t}).
\end{split}%
\end{equation}%
The pump laser owns the driving amplitude $\Omega _{pu}=\sqrt{2P_{pu}\kappa
/\hbar \omega _{pu}}$ and the frequency $\omega _{pu}$. The probe beam has
the driving amplitude $\Omega _{pr}=\sqrt{2P_{pr}\kappa /\hbar \omega _{pr}}$
with the frequency $\omega _{pr}$, where $P_{pu}(P_{pr})$ is the input power
of the pump (probe) field and $\kappa $ is the total decay rate of the left
cavity.

\section{FORECASTS}

We use a silicon nitride microdisk and a silica nanosphere with density of $%
\rho _{1}=2.7$g/cm$^{3}$, $\rho _{2}=2.3$g/cm$^{3}$ and the dielectric
constant $\epsilon _{1}=4$, $\epsilon _{2}=2$, respectively. The mechanical
frequency $\omega _{j}$ depends on the intracavity intensity, it can be
modulated by the power of trapping beam[32]. The cavities are driven with
two trapping lasers of wavelength $\lambda =1.5\mu m$ and power $P_{1}=5.06W$
and $P_{2}=2.87W$ for a disk and sphere, respectively, corresponding to an
axial trap frequency of $\omega _{1}/2\pi =\omega _{2}/2\pi =10$kHz.

The quality factor one would get for levitated resonators are solely
determined by the air molecule impacts. Random collisions with residual air
molecules provide the damping $\gamma _{j}=\omega _{j}Q_{j}^{-1}$ and thus,
the quality factor due to the gas dissipation can be defined as $Q_{1}=\pi
\omega _{2}\rho _{1}\nu d/32p$ for the microdisk and $Q_{2}=\pi \omega
_{2}\nu \rho _{2}r_{2}/16p$ for the nanosphere[24]. Here $\nu =\sqrt{%
k_{B}T/m_{gas}}$ is the thermal velocity of the gas molecules, $p$ the gas
pressure. Let us consider the levitated microdisk with the radius $%
r_{1}=75\mu m$, thickness $d=1\mu m$ and the levitated nanosphere with the
radius $r_{2}=120nm$. We get $Q_{1}=4.3\times 10^{11}$ and $Q_{2}=8.8\times
10^{10}$ for the ultralow pressure $p=10^{-11}$mbar, at room temperature($%
T=300$K).

The coupling strength $\beta $ can be obtained from Eq.(5) by setting
different $n$. In what follows, we take the pump-cavity detuning $\Delta
_{pu}=0$. Here we use the Heisenberg equation of motion to solve the
Hamiltonian of levitation-cavity system. By solving the Heisenberg equation,
we can obtain the transmission of the probe beam, $\left\vert t\right\vert
^{2}$, defined as the ratio of the output and input field amplitudes at the
probe frequency[35] (see the supplementary).

Then we depict the transmission $\left\vert t\right\vert ^{2}$ of the probe
beam as a function of the probe-pump detuning $\delta $ in Fig.2. At first
we assume $\beta =\beta _{casimir}=0$, then we get an enhanced peak which is
located at $\delta =\omega _{1}=10$kHz, just corresponds to the fundamental
frequency of the levitated microdisk as shown by the black curve. Without
the presence of extra dimensions$(n=0)$, we can only consider the
Casimir-Polder coupling $\beta _{casimir}=2.9\times 10^{-7}Hz$ . Then we can
find the resonance peak suffers a splitting in the spectrum characterized by
the blue curve. By applying the extra dimensions based ADD theory, for $n=2$%
, $\beta =1.4\times 10^{-6}Hz$. The resonance frequency splitting can be
amplified significantly in the spectrum as shown by the red curve. Here the
transmitted spectrum of the probe laser can be effectively modulated by the
number of extra dimensions. Without any interaction, one can obtain
significant transmission of the probe laser at the resonant region. When
gravity deviates from $1/r^{2}$ in 4+2 extra dimensions, considering the
Casimir background force, the enhanced peak splits and separates. The linear
increase of $L$ with $\beta $ reminds us of the possibility to detect the
gravity strength between resonators by measuring the separation in the
transmission spectrum. Their relationship can be expressed by $L=2(\beta
+\beta _{casimir})$, which strongly reveals the deviation from gravitational
inverse-square law, namely, a sign indicative of large extra dimensions. The
resolution depends on the full width of half maximum(FWHM) of the peak, thus
the minimal detectable coupling strength $\beta _{\min }=L_{\min }/2=FWHM/2$%
. Considering the peak FWHM in Fig.2 approximates $2\times 10^{-7}$Hz, one
can obtain $\beta _{\min }\approx 0.1\mu Hz.$

The levitated resonators with shorter separation can set better constraints
since the gravitational strength rapid rise with decreasing the displacement
$r_{0}$. However the system with smaller $r_{0}$ will suffer lager Casimir
background force noise which goes up in proportion to the 6th power of $r_{0}
$. We depict Fig.3 to show the coupling-displacement function of the bulk
and Casimir interaction. Then we depict the horizontal dash black line to
show the minimal detectable coupling rate $\beta _{\min }$ which corresponds
to $0.1\mu Hz$. The solid lines represent the detectable space, the dot
lines represent the undetectable region. We can find that a separation of
range of $8\mu m$ is a good compromise between the two limit
factors(separation and noise). In this conditions, the Casimir coupling is
smaller than the gravitational coupling, $\beta _{casimir}\approx \beta /5$,
thus we expect the contribution of the bulk will be able to show itself
clearly on the probe spectrum, performing a test with low Casimir background
noise. Moreover, as shown in Fig.1(a), the middle mirror between two
cavities can be used to minimize the electrostatic and Casimir background
forces by preventing direct coupling between the masses. Thus the shield
mirror can attenuate this Casimir interaction even further, rendering the
effect negligible. This means that we can first fix the trapping frequency
of the microdisk at $\omega _{1}=10kHz$, and adjust the frequency of the
nanosphere approach to $\omega _{1}$ via modulating the trapping power in
the right cavity. If the bulk exist, then we can obtain the vibrational mode
splitting of the microdisk resonator clearly on the probe spectrum, on the
contrary, the resonance peak will fix on the original frequency without the
bulk. In the past few years, the new forces measurement in the Casimir
regime relies on the isoelectronic technique[5,10,13]. This technique will
induce vibration noises between the bimorph and the single-crystal silicon
cantilever, the sensitivity is not able to probe forces below the level of
aN[5]. Compare to the previous experiment, the sensitivity can be improved
by the using of a specific geometry(sphere-plane interaction) to generate
the Casimir-Polder potential and the shield mirror to suppress the effects
of the Casimir background. Then the pump-probe technology can be used to
read the weak frequency splitting. We list all the main optomechanical
parameters in the Table I.
\begin{table}[tbp]
\begin{ruledtabular}
\caption{Optomechanical parameters of the levitated microdisk and nanosphere}
\begin{tabular}{lllll}
Parameter ~ & Units ~ & Value ~ \\ \hline
Separation distance $r0$ & $\mu m$ & $8$\\
Nanosphere radius $r_{2}$ & $nm$ & $120$\\
Microdisk radius $r_{1}$ & $\mu m$ & $75$\\
Microdisk thickness $d$ & $\mu m$ & $1$\\
Microdisk frequency $\omega _{1}$ & $kHz$ & $10$\\
Nanosphere frequency $\omega _{2}$ & $kHz$ & $10$\\
Trapping wavelength $\lambda$ & $\mu m$ & $1.5$\\
Pump driving amplitude $\Omega _{pu}$ & $GHz$ & $0.1$\\
Probe driving amplitude $\Omega _{pr}$ & $MHz$ & $1$\\
Total cavity decay $\kappa$ & $GHz$ & $0.1$\\
Air pressure $p$ & $mbar$ & $10^{-11}$\\
Room temperature $T$ & $K$ & $300$\\
Optomechanical coupling rate $g_{1}$ & $Hz$ & $7.5$\\
Pump-cavity detuning $\Delta _{c}$ & $Hz$ & $0$\\
\end{tabular}
\end{ruledtabular}
\end{table}

The mechanism underlying these effects can be explained as four-wave mixing (FWM) in a three
energy levels system. The simultaneous presence of a pump field and a probe
field generates a radiation pressure force at the beat frequency, which
drives the motion of the oscillator near its resonance frequency.In
Fig.1(b), we let $\left\vert N\right\rangle $, $\left\vert
n_{1}\right\rangle $ and $\left\vert n_{2}\right\rangle $ denote the number
states of the cavity photon, microdisk phonons, and nanosphere phonons
respectively. $\left\vert N,n_{1},n_{2}\right\rangle \leftrightarrow
\left\vert N+1,n_{1},n_{2}\right\rangle $ transition changes the cavity
field, $\left\vert N+1,n_{1},n_{2}\right\rangle \leftrightarrow \left\vert
N,n_{1}+1,n_{2}\right\rangle $ transition is caused by the radiation
pressure coupling. In the system,the coupling between two resonators adds a
fourth level which can be mainly modified by the gravitational interaction.
The coupling breaks down the symmetry of the OMIT interference, the single
OMIT transparency window is split into two transparency windows, which
yields the quantum coupling induced NMS as shown in Fig.2.
\begin{figure}[tbp]
\includegraphics[width=8.5cm]{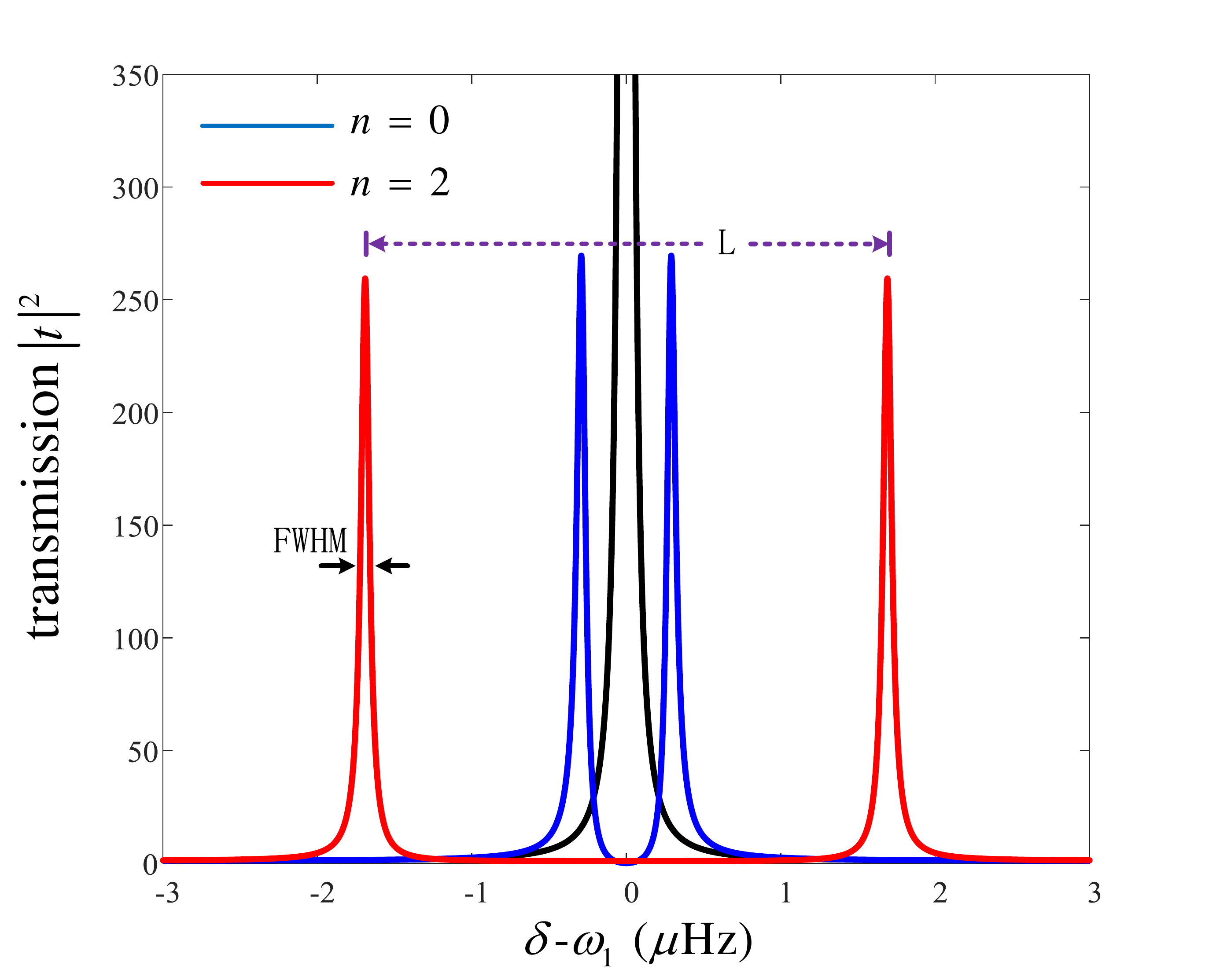}
\caption{The $n=2$ extra dimensions modulate transmission spectrum of the
probe field. A peak splitting caused by gravitational deviation and Casimir
coupling can be well recognized in the spectrum. In this conditions, the
Casimir coupling is about 5 times less than the gravitational coupling}
\end{figure}
\begin{figure}[tbp]
\includegraphics[width=8cm]{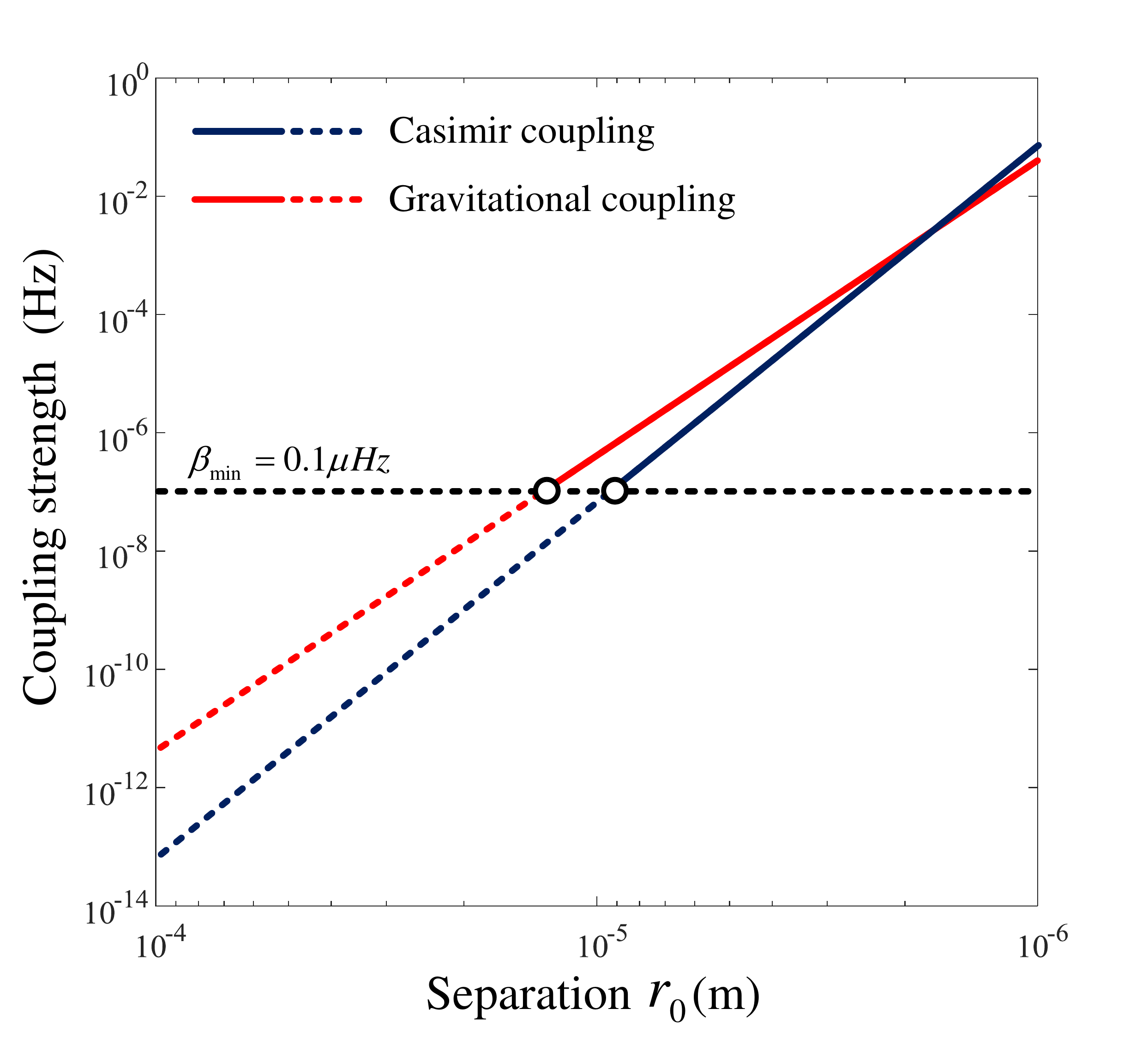}
\caption{The coupling-displacement function of bulk induced coupling and
Casimir-Polder coupling. The horizontal dash black line shows the minimal
detectable coupling strength $\protect\beta _{\min }$. }
\end{figure}
\ \ \ \

\section{CONSTRAINTS AND LIMITS}

The spectral resolution depends on the full width at half maximum(FWHM) of the oscillation peak. The
result shows that smaller linewidth can be achieved by increasing the Q
factor of the levitated resonators.
\begin{figure}[tbp]
\includegraphics[width=8cm]{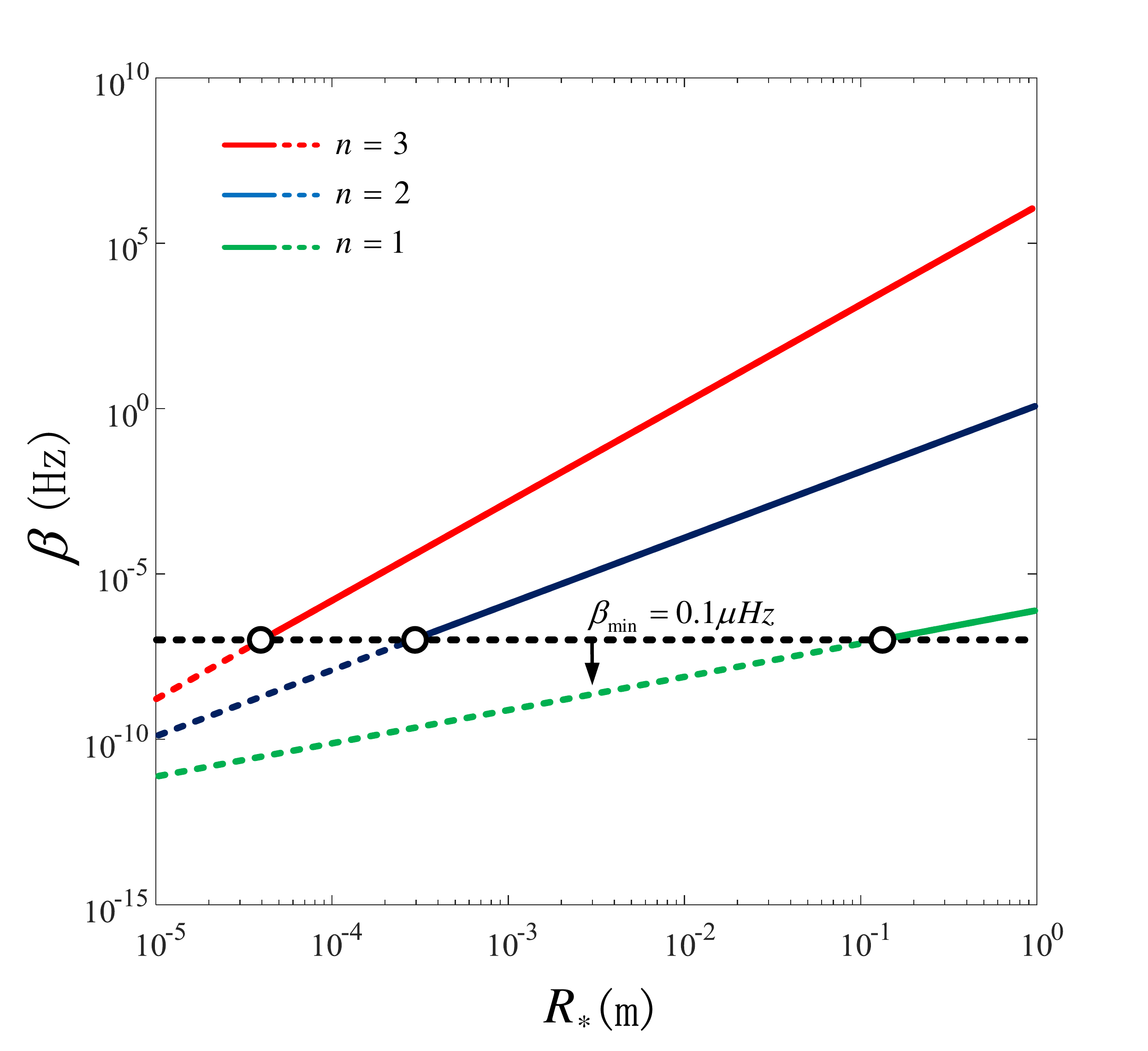}
\caption{Limit on the compactification range $R_{\ast }$ and the bulk
induced coupling $\protect\beta $ with the ADD formalism. The parameters
take the same values as in Table.I.}
\end{figure}
Fig.4 presents the constraints on $R_{\ast }$ for the number of extra
dimensions, $n=1,2,3$.The sloping lines represent the calculated coupling
rate $\beta $ for different $n$ and $R_{\ast }$.Then we depict the
horizontal dash black line to indicate the limits of detectable coupling
strength $\beta _{\min }$. $R_{\ast }$ and $\beta $ are constrained to be
larger than the values of the intersections in the picture. The solid lines
in the figure represent the detectable parameter space, the dot lines
represent the undetectable region. The arrow indicates that the sensitivity
can be improved through optomechanical oscillators with higher Q factors.
Considering the minimum measurable gravitational coupling rate $\beta _{\min
}=0.1\mu Hz$, we get the precision for the forces measurement as $F_{\min
}=2.2\times 10^{-4}aN$. The Q factor of the optically trapped particles is
limited only by collisions with residual air molecules, thus the sensitivity
can be improved by decreasing the air pressure. The lower pressure limit of
sputter-ion pumps is in the range of $10^{-11}$mbar. Lower pressures in the
range of $10^{-12}$ mbar can only be achieved when the sputter-ion pump
works in a combination with other pumping methods[36,37]. In our
considerations, a conservative value is taken $p=10^{-11}$mbar , which is
usually required for achieving ultrahigh-Q mechanical oscillators and the
ultrasensitive measurements in the levitated optomechanical
system[12,19,23,24,38]. If we choose a lower air pressure ($p=10^{-12}$mbar)
,this scheme will gain a sensitivity of $F_{\min }=4.0\times 10^{-5}aN$. Our
scheme yields a 4-5 orders of magnitude improvement for the force sensing in
the micro-scale[5].

The fundamental fluctuation processes impose the ultimate limits upon the
sensitivity of the detection. Frequency stability $\Delta \omega _{\min }$
is key to performance of micro-resonators and their applications in the
thermomechanical noise system. For the case of the high quality factor $Q\gg
1$, the minimum detectable frequency shift limited by thermomechanical
fluctuations of the levitated microdisks can be estimated via[39],%
\begin{equation}
\Delta \omega _{\min }\approx \sqrt{\frac{k_{B}T\omega _{1}\Delta f}{%
E_{c}Q_{1}}}.
\end{equation}%
where, $\Delta f=1/2\pi \tau $ the measurement bandwidth which is dependent
upon the measurement averaging time $\tau $. $E_{c}=m_{1}\omega
_{1}^{2}\langle x_{c}^{2}\rangle $ represents the maximum drive energy, $%
\left\langle x_{c}\right\rangle $ is the maximum root mean square(rms) level
produced a predominantly linear response. For a Gaussian field distribution,
the nonlinear coefficients are given by $\xi =-2/W^{2}$[40], here $W\approx
75\mu m$ is the trapping beam waist. For the displacements $\left\vert
x_{c}\right\vert \ll \left\vert \xi \right\vert ^{-1/2}=5\times 10^{-5}m$
the nonlinearity is negligible. In our considerations, $x_{c}$ is taken to
be 2 orders of magnitude smaller, we choose $x_{c}\approx 10^{-7}m$.
Considering the quality factor of the microdisk $Q_{1}=4.3\times 10^{11}$,
we can obtain $\Delta \omega _{\min }=4.5\times 10^{-8}Hz<\beta _{\min }$
for the measurement bandwidth $\Delta f=10^{-3}Hz$ at the room temperature.
It is corresponding to the measurement time $\tau =160s$. The result shows
that the thermomechanical noise can be controlled at a level lower than the
claimed sensitivity.

\section{CONCLUSION}

We have analyzed the large extra dimensions induced coupling between two
levitated resonators with quantum optomechanics. Our study reports a design
for probing gravitational deviation in the range of 8$\mu m$ via the
pump-probe optical technology in cavity. The transparency peak will suffer a
distinct splitting and get apart with the non-Newtonian gravity being taken
into consideration. The gravitational strength, characterized by the
coupling strength $\beta $, can also be determined by the splitting distance
$L$. The Casimir-Polder coupling rate is about 5-fold smaller than the
gravitational strength, and can be attenuated even further by the shield
mirror. It allows the precision measurement in the Casimir regime without
the isoelectronic technique. We hope that the precision can be significantly
enhanced by experiments in ultrahigh vacuum.

\begin{acknowledgments}
This work was supported by the National Natural Science Foundation of China
(Nos.11274230 and 11574206), the Basic Research Program of the Committee of
Science and Technology of Shanghai (No.14JC1491700).
\end{acknowledgments}

\end{document}

\preprint{APS/123-QED}

\title{Supplementary Information}% Force line breaks with \\
\thanks{}%

\author{Jian Liu}
 \altaffiliation[]{}%Lines break automatically or can be forced with \\
\author{Ka-Di Zhu}%
 \email{zhukadi@sjtu.edu.cn}
\affiliation{Key Laboratory of Artificial Structures and Quantum Control (Ministry of
Education), School of Physics and Astronomy, Shanghai Jiao Tong
University, 800 DongChuan Road, Shanghai 200240, China,
Collaborative Innovation Center of Advanced Microstructures, Nanjing, China
}%

\date{\today}% It is always \today, today,
             %  but any date may be explicitly specified

\begin{abstract}

\keywords{}

\end{abstract}

\pacs{Valid PACS appear here}% PACS, the Physics and Astronomy
                             % Classification Scheme.
%\keywords{Suggested keywords}%Use showkeys class option if keyword
                              %display desired
\maketitle

%\tableofcontents

\section{The Heisenberg equations}

The Hamiltonian of the system in a rotating frame at the pump frequency $%
\omega _{pu}$ and the probe pulse frequency $\omega _{pr}$ reads as follow:%
\begin{equation}
\begin{split}
H& =\hbar \omega _{c}c^{+}c+\hbar \omega _{1}a_{1}^{+}a_{1}+\hbar \omega
_{2}a_{2}^{+}a_{2}-\hbar gc^{+}c(a_{1}^{+}+a_{1}) \\
& +\hbar (\beta +\beta _{casimir})\cdot (a_{1}^{+}a_{2}+a_{1}a_{2}^{+}) \\
& -i\hbar \Omega _{pu}(c-c^{+})-i\hbar \Omega _{pr}(ce^{i\delta
t}-c^{+}e^{-i\delta t}).
\end{split}%
\end{equation}

We define the operator $S_{j}=a_{j}^{+}+a_{j}$ , the detuning of the probe
and pump field $\delta =\omega _{pu}-\omega _{pr}$ and the pump-cavity
detuning $\Delta _{pu}=\omega _{c}-\omega _{pu}$. In what follows, we deal
with the mean response of the system to the probe field in the presence of
the coupling, let $\left\langle c\right\rangle $, $\left\langle
c^{+}\right\rangle $ and $\left\langle S_{1,2}\right\rangle $ be the
expectation values of operators $c$, $c^{+}$ and $S_{1,2}$, respectively.
According to the Heisenberg equation of motion and the commutation relations
$[c,c^{+}]=1$, $[a,a^{+}]=1.$ In a frame rotating with the frequency of the
pump field, the temporal evolutions of $c$ and $S_{1,2}$ can be obtained and
the corresponding equations are given by adding the damping terms%
\begin{equation}
\begin{split}
& \frac{d\left\langle c\right\rangle }{dt}=-(i\Delta _{pu}+\kappa
)\left\langle c\right\rangle +ig_{1}\left\langle S_{1}\right\rangle
\left\langle c\right\rangle \\
& +\Omega _{pu}+\Omega _{pr}e^{-i\delta t},
\end{split}%
\end{equation}%
\begin{equation}
\begin{split}
& \frac{d^{2}\left\langle S_{1}\right\rangle }{dt^{2}}+\gamma _{1}\frac{%
d\left\langle S_{1}\right\rangle }{dt}+(\omega _{1}^{2}+\beta
^{2})\left\langle S_{1}\right\rangle -\beta (\omega _{1}+\omega
_{2})\left\langle S_{2}\right\rangle \\
& =2g_{1}\omega _{1}\left\langle c^{+}\right\rangle \left\langle
c\right\rangle ,
\end{split}%
\end{equation}%
\begin{equation}
\begin{split}
& \frac{d^{2}\left\langle S_{2}\right\rangle }{dt^{2}}+\gamma _{2}\frac{%
d\left\langle S_{2}\right\rangle }{dt}+(\omega _{2}^{2}+\beta
^{2})\left\langle S_{2}\right\rangle -\beta (\omega _{1}+\omega
_{2})\left\langle S_{1}\right\rangle \\
& =2g_{1}\beta \left\langle c^{+}\right\rangle \left\langle c\right\rangle .
\end{split}%
\end{equation}%
To solve these equations, we make the ansatz as follows:
\begin{equation}
\left\langle c(t)\right\rangle =c_{0}+c_{+}e^{-i\delta t}+c_{-}e^{i\delta t},
\end{equation}%
\begin{equation}
\left\langle S_{1}(t)\right\rangle =S_{10}+S_{1+}e^{-i\delta
t}+S_{1-}e^{i\delta t},
\end{equation}%
\begin{equation}
\left\langle S_{2}(t)\right\rangle =S_{20}+S_{2+}e^{-i\delta
t}+S_{2-}e^{i\delta t}.
\end{equation}%
Substituting Eqs.(5)-(7) into Eqs.(2)-(3), respectively, equating terms
with the same time dependence, and working to the lowest order in $\Omega
_{pr}$ but to all orders in $\Omega _{pu}$, we can obtain
\begin{equation}
c_{0}=\frac{\Omega _{pu}}{i\Delta _{pu}+\kappa -i\delta -ig_{1}S_{10}},
\end{equation}%
\begin{equation}
c_{+}=\frac{\Omega _{pr}+ic_{0}g_{1}S_{1+}}{i\Delta _{pu}+\kappa -i\delta
-ig_{1}S_{10}},
\end{equation}%
\begin{equation}
c_{-}=\frac{ic_{0}g_{1}S_{1-}}{i\Delta _{pu}+\kappa +i\delta -ig_{1}S_{10}},
\end{equation}%
and
\begin{equation}
(\omega _{1}^{2}+\beta ^{2})S_{10}-\beta (\omega _{1}+\omega
_{2})S_{20}=2g_{1}\omega _{1}c_{0}^{2},
\end{equation}%
\begin{equation}
S_{1+}=\frac{\beta (\omega _{1}+\omega _{2})S_{2+}+2g_{1}\omega
_{1}(c_{0}^{\ast }c_{+}+c_{0}c_{-}^{\ast })}{-\delta ^{2}-i\gamma _{1}\delta
+\omega _{1}^{2}+\beta ^{2}},
\end{equation}%
\begin{equation}
S_{1-}=\frac{\beta (\omega _{1}+\omega _{2})S_{2-}+2g_{1}\omega
_{1}(c_{0}^{\ast }c_{-}+c_{0}c_{+}^{\ast })}{-\delta ^{2}+i\gamma _{1}\delta
+\omega _{1}^{2}+\beta ^{2}},
\end{equation}

\begin{equation}
(\omega _{2}^{2}+\beta ^{2})S_{20}-\beta (\omega _{1}+\omega
_{2})S_{10}=2g_{1}\beta c_{0}^{2},
\end{equation}%
\begin{equation}
S_{2+}=\frac{\beta (\omega _{1}+\omega _{2})S_{1+}+2g_{1}\beta (c_{0}^{\ast
}c_{+}+c_{0}c_{-}^{\ast })}{-\delta ^{2}-i\gamma _{2}\delta +\omega
_{2}^{2}+\beta ^{2}},
\end{equation}%
\begin{equation}
S_{2-}=\frac{\beta (\omega _{1}+\omega _{2})S_{1-}+2g_{1}\beta (c_{0}^{\ast
}c_{-}+c_{0}c_{+}^{\ast })}{-\delta ^{2}+i\gamma _{2}\delta +\omega
_{2}^{2}+\beta ^{2}}.
\end{equation}%
Solving Eqs. (8)-(16), we obtain in the steady state,
\begin{equation}
c_{+}=\frac{\Omega _{pr}[(A_{1}A_{2}-W^{2})(U-V)+Y\omega _{0}]}{%
(A_{1}A_{2}-W^{2})(U^{2}-V^{2})+2VY\omega _{0}},
\end{equation}%
and
\begin{equation}
\Omega _{pu}^{2}=[\kappa ^{2}+(\Delta _{pu}-g_{1}S_{10})^{2}]\omega _{0},
\end{equation}%
with $\omega _{0}=\left\vert c_{0}\right\vert ^{2}$, $U=\kappa -i\delta $, $%
V=i\Delta _{pu}-ig_{1}S_{10}$, $A_{j}=-\delta ^{2}-i\gamma _{j}\delta
+\omega _{j}^{2}+\beta ^{2}$($j=1,2$), $W=\beta (\omega _{1}+\omega _{2})$, $%
R_{1}=2g_{1}\omega _{1}$, $R_{2}=-2g_{1}\beta $, $%
Y=ig_{1}(A_{2}R_{1}+WR_{2}) $. Then $S_{10}$, $S_{20}$ can be resolved by
\begin{equation}
(\omega _{1}^{2}+\beta ^{2})S_{10}-WS_{20}=R_{1}\omega _{0},
\end{equation}%
\begin{equation}
(\omega _{2}^{2}+\beta ^{2})S_{20}-WS_{10}=R_{2}\omega _{0}.
\end{equation}%
The transmission of the probe beam, defined as the ratio of the output and
input field amplitudes at the probe frequency is then given by using the input-output relation[1],
\begin{equation}
t(\omega _{pr})=\frac{\Omega _{pr}/\sqrt{2\kappa }-\sqrt{2\kappa }c_{+}}{%
\Omega _{pr}/\sqrt{2\kappa }}=1-\frac{2\kappa c_{+}}{\Omega _{pr}}.
\end{equation}